\documentclass[preprint2]{aastex}

\slugcomment{Not to appear in Nonlearned J., 45.}

\shorttitle{}
\shortauthors{van der Walt}

\begin{document}


\title{A near-infrared study of the star forming region RCW 34}


\author{D.J. van der Walt, H.M. de Villiers, R.J.Czanik}
\affil{Centre for Space Research, North-West University, Potchefstroom,
  South Africa}

\begin{abstract}
We report the results of a near-infrared imaging study of a $7.8 \times 7.8$
arcmin$^2$ region centered on the 6.7 GHz methanol maser associated with the RCW
34 star forming region using the 1.4m IRSF telescope at Sutherland. A total of
1283 objects were detected simultaneously in J, H, and K for an exposure time of
10800 seconds. The J-H, H-K two-colour diagram revealed a strong concentration
of more than 700 objects with colours similar to what is expected of reddened
classical T Tauri stars. The distribution of the objects on the K {\it vs} J-K
colour-magnitude diagram is also suggestive that a significant fraction of the
1283 objects is lower mass pre-main sequence stars. We also present the
luminosity function for the subset of about 700 pre-main sequence stars and show
that it suggests ongoing star formation activity for about $10^7$ years. An
examination of the spatial distribution of the pre-main sequence stars shows
that the fainter (older) part of the population is more dispersed over the
observed region and the brighter (younger) subset is more concentrated around
the position of the O8.5V star. This suggests that the physical effects of the
O8.5V star and the two early B-type stars on the remainder of the cloud out of
which they formed, could have played a role in the onset of the more recent
episode of star formation in RCW 34.
\end{abstract}


\keywords{stars: formation --- stars: pre-main sequence --- ISM: individual
  (RCW 34)}

\section{Introduction}
Studies on star formation can very broadly be classified as either the study of
the process of star formation as such or as the study of the larger scale
processes that trigger star formation in molecular clouds and which determines
the mode of star formation, ie. clustered or distributed star
formation. Although large strides have been made over the last couple of decades
in understanding the process of star formation, an understanding of what
determines the mode of star formation, which is equally important to
understanding the process of the formation of individual stars, is still
lacking. The two aspects are, obviously, not independent of each other and
together play a role in determining important properties such as, for example,
the stellar initial mass function and the star formation efficiency. To achieve
the goal of a theoretical understanding of the physical process determining the
mode of star formation, empirical information on a sufficiently large sample of
embedded young clusters is necessary \citep{lada2010}. The present near-infrared
study of RCW 34 attemps to add to this effort.

RCW 34 is a southern high mass star forming region that has been studied at
irregular intervals over a number of decades. The number of observational
studies that was directly aimed at investigating RCW 34 is quite small. We
review those that are relevant to the present study. RCW 34 was first catalogued
by \citet{rodgers1960} as one of the southern $\mathrm{H\alpha}$ emission
regions with a diameter less than 2 $\times$ 2 arcminutes. The exciting star of
RCW 34 also appears in the catalogue of stars in reflection nebulae compiled by
\citet{vandenbergh1975}. Based on this catalogue the exciting star in RCW 34 is
also refered to in the literature as either Vela R2 25a or vdBH 25a. Following
up on the catalogue of \citet{vandenbergh1975}, \citet{herbst1975a} did UBV and
MK spectroscopy on the stars in the catalogue and classified the exciting star
in RCW 34 as of spectral type O9. In a further analysis of the UBV data,
\citet{herbst1975b} did main-sequence fitting on the stars of Vela R2
association. This resulted in a distance of about 870 pc to the
association. However, the exciting star of RCW 34 was found to lie well below
the ZAMS, which lead \citet{herbst1975b} to consider it to be
peculiar. \citet{herbst1975b} explained the position of RCW 34 well below the
ZAMS as due to it being either an O star located further than the Vela R2
association or an O-type pre-main sequence star with a ``grey'' circumstellar
shell associated with Vela R2.  \citet{vittone1987} was the first to follow up
on the finding of \citet{herbst1975b} that the exciting star of RCW 34 lie well
below the ZAMS for the Vela R2 association. These authors also concluded that
vdBH 25a is an O9 type star but that the distance to the star is uncertain.

The first really comprehensive study of RCW 34 aimed to answer the questions on
the nature of vdBH 25a and its distance is that of \citet{hm1988}. Using a
variety of observations and data from the literature, \citet{hm1988} concluded
that RCW 34 is located at 2.9 kpc, well beyond the Vela R2 association, and that
its exciting star is a heavily reddened (A$_v$ = 4.2) O8.5V ZAMS star. There is
therefore nothing peculiar about the exciting star of RCW 34. \citet{hm1988}
also interestingly remarks that the presence of an $\mathrm{H_2O}$ maser might
signify ongoing star formation in RCW 34.

The most recent study of RCW 34 is that of \citet{bik2010}. These authors made
near-infrared H- and K-band observations on a $104^{\prime\prime} \times
60^{\prime\prime}$ field centered on the O8.5V ionizing star, using the Integral
Field instrument SINFONI on UT4 of the VLT at Paranal, Chile. In addition they
also used archival data obtained with IRAC \citep{fazio2004} on board the {\it
  Spitzer} satellite. Using this data \citet{bik2010} identified three regions
associated with RCW 34: The first is a large bubble detected in the IRAC images
in which a cluster of intermediate- and low-mass class II objects is found. The
second is the \ion{H}{2} region located at the northern edge of the bubble and
which is ionized by three OB stars. The third region is a photon-dominated
region north of the \ion{H}{2} region and which marks the edge of a dense
molecular cloud traced by H$_2$ emission. Several class 0/I objects associated
with this cloud were also identified indicating recent star formation
activity. \citet{bik2010} revised the distance to RCW 34 to 2.5 $\pm$ 0.2 kpc
and derived an age estimate of 2 $\pm$ 1 Myr from the properties of the pre-main
sequence stars inside the \ion{H}{2} region.

The observations presented here differ from earlier near-infrared (NIR)
observations, and in particular from that of \citet{bik2010} in that we imaged a
7.8$^\prime$ $\times$ 7.8$^\prime$ region centered on the 6.7 GHz methanol maser
located at coordinates $\mathrm{\alpha(2000) = 08^h56^m24^s.9}$ and
\\ $\mathrm{\delta(2000) = -43^{\circ}05^{\prime}40^{\prime\prime}.4}$. This is
significantly larger than the 104$^{\prime \prime}$ $\times$ 60$^{\prime
  \prime}$ region covered by \citet{bik2010}. This much larger field gives us
the opportunity to also investigate possible star formation activity especially
to the north of the shocked region where the \ion{H}{2} region interacts with
the molecular cloud and created a photo-dissociation region \citep{bik2010}.
Except for the class II methanol maser which indicates the presence of a very
young high mass star there appears to be no other signs of current high mass
star formation activity to the north of the \ion{H}{2} region. This does not
exclude the presence of lower mass stars, however.

Within the framework of a broader view on star formation as outlined above, our
results should be seen as complimentary to that of eg. \citet{bik2010} and
others which can help to understand the star formation history of RCW 34. Our
aim here is to present some basic properties of the embedded cluster associated
with RCW 34 as derived from our NIR observations. 

\section{Observations, data reduction, and calibration} 

The observations used in this study can be regarded as archival data in the
sense that the data were not collected by the authors themselves for the primary
purpose of studying RCW 34 as is done here. The data used here was collected by
GG Nyambuya in 2005, however, for a completely different purpose. This caused
the data to be lacking in certain aspects, for example in that there were no
control field observations.

The observations were made on the nights of 2005, April 18, 22 and 25 with the
SIRIUS (Simultaneous-three-colour Infrared Imager for Unbiased Survey) camera of
the Infra Red Survey Facility's 1.4m telescope located at the SAAO's Sutherland
observatory. SIRIUS has three science-grade HAWAII 1024 $\times$ 1024 arrays
with two dicroic mirrors which enable {\em simultaneous} observations in $J$
($\lambda = 1.25 \mu m)$, $H$ ($\lambda = 1.65 \mu m)$ and $K_s$ ($\lambda =
2.25 \mu m)$. Precise colour information can thus be obtained with such
simultaneous observations. The atmospheric conditions for the three nights gave
a FWHM for the seeing disk of 1.1 to 1.4 arcsec for images in the $K_s$
band. The field of view was 7.8 $\times$ 7.8 arcmin$^2$ with a pixel scale of
0.45 arcsec per pixel.

The total exposure of the observations consisted of three nights of 12 dithered
sets per night, each set consisting of 10 ditherings of 30s each. This resulted
in a total exposure time of 10800 seconds for the three nights. The initial data
reduction was done at the University of Cape Town using the SIRIUS pipe-line
software which include all the basic reduction steps such as dark subtraction,
flat fielding, and sky subtraction to produce a single image for one night's
observations. The images of the three nights were median combined to produce a
final image.

The IRAF task DAOFIND was used to select stars from the field.  In total 1283
objects were detected in all three bands. The instrumental magnitudes were
calibrated to the 2MASS system. This was done by selecting 40 stars in the
observed field detected in J, H, and K for both the IRSF and 2MASS and for which
the world coordinates of both sets agreed such that the cross identification
between the IRSF and 2MASS is unique. The correlation between IRSF instrumental
and 2MASS apparent magnitudes as well as between IRSF colours and 2MASS colours
of the 40 stars were investigated. This led to the following equations
transforming IRSF instrumental magnitudes to 2MASS apparent magnitudes: \small
\begin{eqnarray*}
\mathrm{J_{2mass}} & = & \mathrm{J_{IRSF} + 0.1175(J-H)_{IRSF}} - 4.2318 \\
\mathrm{H_{2mass}} & = & \mathrm{H_{IRSF} - 0.0761(H-Ks)_{IRSF}} - 4.0974 \\
\mathrm{Ks_{2mass}} & = & \mathrm{Ks_{IRSF} + 0.0240(H-Ks)_{IRSF}} - 5.0834
\end{eqnarray*}
\normalsize
Following this calibration our observations reached J = 19.79, H =
18.43, and K$_s$ = 17.64. 

It is also necessary to comment on adopting a distance of 2.5 kpc to RCW 34
before presenting the results. It has already been noted that \citet{hm1988}
asigned a distance of 2.9 kpc to RCW 34 also using the spectral type of the
exciting star. \citet{bik2010} pointed out that the difference between their
distance estimate and that of \citet{hm1988} lies in the different absolute
magnitudes used. Two kinematic distance estimates for RCW 34 are also
available. The water maser associated with RCW 34 has a velocity
$\mathrm{V_{lsr}=10.8~km\,s^{-1}}$ \citep{braz1982}. Using the rotation curve of
\citet{wouterloot1989} this gives a distance of 2.1 kpc. The second kinematic
distance comes from the velocity of the radio recombination line associated with
the \ion{H}{2} region. \citet{caswell1987} gives this as
$\mathrm{12~km\,s^{-1}}$ which translates to a distance of 2.3 kpc, again using
the rotation curve of \citet{wouterloot1989}. Although a detailed analysis of
the uncertainties of each of the four estimates can be made to decide which to
use, it seems more appropriate for the present to use the mean of the four
estimates which is 2.45 kpc. It therefore seems reasonable to simply use the
distance estimate of \citet{bik2010}, ie. 2.5 kpc.

\section{Results and analysis}

\subsection{The two-colour diagram}

\begin{figure}[h]
   \centering \includegraphics[width=7cm,angle=0,clip]{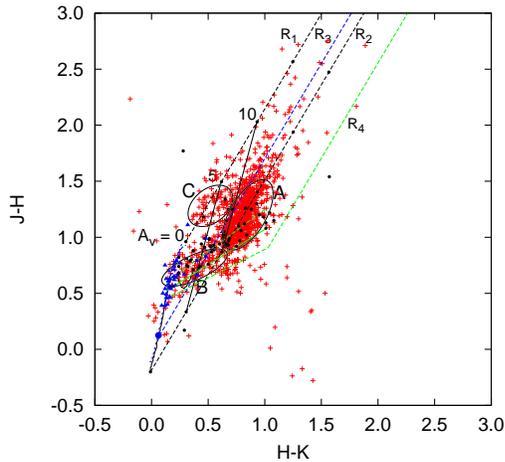}
      \caption{Near-infrared two-colour diagram for 1283 objects (red
        crosses) detected in J, H, and K. Blue triangles are WTTS from
        \citet{strom1989} and black dots CTTS from \citet{strom1989} and
        \citet{cieza2005}.}
\label{fig:rcw34mp}
   \end{figure}

The NIR J-H, H-K two colour diagram has been used in the past by numerous
authors to present NIR photometry of young stellar populations and to identify
different types of objects.  Here we follow the same practice.  The main result
of the present study is presented in Fig. \ref{fig:rcw34mp}. The main sequence
and giant branch were taken from \citet{koornneef1983} and transformed to the
2MASS system. Four reddening lines using the extinction law of \citet{rieke1985}
are also shown. $\mathrm{R_1}$ and $\mathrm{R_2}$ are the reddening lines that
bracket the main sequence and giant branch while $\mathrm{R_3}$ is the reddening
line for a star of spectral type F0, which is more or less the upper mass limit for T
Tauri stars. Weak line T Tauri stars (WTTS) have intrinsic NIR colours
consistent with those of normal dwarf stars \citep{lada1992,meyer1997} and
should therefore fall between reddening lines R$_1$ and R$_3$. The solid green
line gives the classical T Tauri star (CTTS) locus from \citet{meyer1997} and
the dashed green lines the upper and lower boundaries as calculated from the
errors given by these authors. The meaning of reddening line R$_4$ is quite
obvious.

We broadly identify three groupings or clusterings of objects on the two-colour
diagram, with the area that each more or less cover indicated by the ellipses in
Fig. \ref{fig:rcw34mp}. Group A is a strong clustering of objects extending
along and on both sides of reddening line R$_2$ just above its intersection with
the CTTS locus and is the most obvious aspect of the distribution of the 1283
objects on the two-colour diagram. Above reddening line R$_2$ most of the group
A objects lie between reddening lines R$_2$ and R$_3$. There also seems to be a
band of objects starting from the A$_V$ = 10 point on reddening line R$_2$ and
extending almost along the CTTS locus to the lower left with most of the
objects lying inside the boundaries of the CTTS region although some lie just
above the upper boundary. This is group B. Group C is a very loose group of
objects that lie to the left of group A. The reason for identifying group C is
mainly because there seems to be a gap between the bluer objects in group B and
the objects in group C.

It is also seen that there are highly reddended objects with visual extinctions
around 20 as well as objects with very large infrared excesses. For the present
we focus mainly on trying to understand group A which is the most obvious
feature of the distribution on the two colour diagram.

   \begin{figure}[h]
   \centering \includegraphics[width=7cm,angle=0,clip]{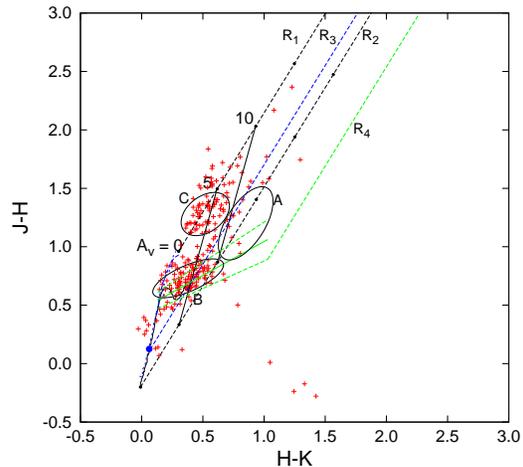}
      \caption{Near-infrared two-colour diagram for IRSF objects
        brighter than J = 16.21 magnitude.}
\label{fig:rcw34_2m}
   \end{figure}

Given the distribution of objects on the two-colour diagram in
Fig. \ref{fig:rcw34mp} the question might be asked as to the reality of the
clustering and the nature of the objects belonging to group A. Although in NIR
studies of other star forming regions it is not uncommon to find objects in the
same region on the two-colour diagram as that of group A, the large number of
objects in the case of RCW 34, is uncommon. The fact that the SIRIUS camera has
three independent arrays means that the detection of an object in all three the
NIR bands implies a real detection and not some artifact on the array. The
existence of a large number of objects with NIR colours as reflected by the
group A objects should therefore be regarded as real. Since we have no
spectroscopic identification of any of the objects it is necessary to proceed
without such information.

Focusing attention on the group A objects, we note the following. First, they
cluster just above the CTTS locus. In fact it would seem as if the CTTS locus
acts as an approximate lower boundary for these objects on the two-colour
diagram. Second, a significant fraction of the objects lie to the right of
reddening line R$_2$. These objects have an infrared excess and cannot be
dereddended to the main sequence. It is also seen that a further significant
fraction of the remainder of the group A objects lie above the CTTS locus and
between reddening lines R$_2$ and R$_3$. Although these objects can be
dereddened to the main sequence, such a dereddening would imply the presence of
a very large number of stars of spectral type earlier than F0. Although the
visual extinction of these stars would be between 10 and 15 magnitudes, the
dereddening would also result in some being ionizing stars which definitely
would have been detected as \ion{H}{2} regions in radio continuum surveys or as
bright infrared sources. Only three OB stars are, however, associated with RCW
34 \citep{bik2010}. 

We note that in some studies similar to the present one, such as eg. that of
\citet{dahm2005} and \citet{barentsen2011}, use has been made of 2MASS for the
NIR photometry. Inspection of the magnitudes of the 2MASS objects in the same
field covered by the present IRSF observations showed that the limiting J-band
magnitude is 16.12 for objects detected simultaneously in J, H, and K. Figure
\ref{fig:rcw34_2m} shows the two-colour diagram for IRSF detections with J
magnitude brighter than 16.12.  What is important to note is the near absence of
objects in the group A region. On the other hand, it is seen that there still are
quite a number of objects in group B. This suggests that the majority of the
group A objects is a population of fainter objects revealed by our significantly
deeper imaging of RCW 34 compared to 2MASS. As already argued, it seems unlikely
that the group A objects are simply reddened main-sequence stars. To avoid a
completely skewed IMF they rather seem to be lower mass objects associated with
the RCW 34 star forming region.

   \begin{figure}[ht]
   \centering \includegraphics[width=7cm,angle=0,clip]{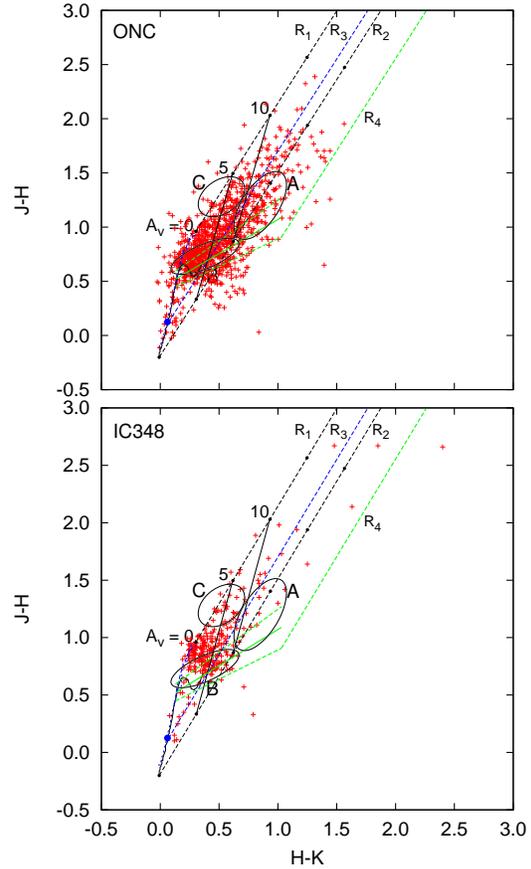}
      \caption{Near-infrared two-colour diagrams for the Orion Nebula Cluster
        (top panel) and IC 348 (lower panel).}
\label{fig:orionic348}
   \end{figure}

The fact that the group A objects cluster above the CTTS locus is suggestive
that they might be reddended CTTS.  Some early NIR imaging of embedded clusters
such as eg. IC348 \citep{ladaea1995}, NGC1333 \citep{lada1996}, L1630
\citep{li1997}, as well as some more recent imaging eg. in the cases of NGC 2316
\citep{teixeira2004} and the Horsehead Nebula \citep{bowler2009} do not show
similar large numbers or even has a lack of apparent CTTS's. This may well be due to
these surveys not being deep enough or that intrinsically there is an absence of
large numbers of CTTS in these star forming regions. Although not in such large
numbers as in the case of RCW 34, other star forming regions such as eg. NGC
7538 \citep{balog2004} do show a significant number of cluster members which, in
terms of Fig. \ref{fig:rcw34mp}, lie just below reddening line R$_2$ and thus can
not be dereddened to the main-sequence as is also the case for RCW
34. Similarly, for AFGL5180 \citep{devine2008} quite a significant fraction of
the objects on the two-colour diagram lie below the reddening line R$_2$ but
above the T Tauri locus of \citet{meyer1997}. 

In all of the above mentioned studies objects lying to the right of reddening
line R$_2$ and above the T Tauri locus are regarded as pre-main sequence
stars. In fact, \citet{lada1992} have shown that pre-main sequence stars
classified as CTTS based on the equivalent widths of their H$\alpha$ emission
lie in the group A region. Other studies eg. that of \citet{strom1989},
\citet{cieza2005}, \citet{luhman1998}, and \citet{barentsen2011} confirm
this. In Fig. \ref{fig:rcw34mp} we also show the positions of the
spectroscopically identified CTTS from \citet{strom1989} and \citet{cieza2005}
as well as the WTTS from \citet{strom1989}. It is seen that the
spectroscopically identified CTTS are associated with the group A and B
regions. \citet{luhman1998} also found that for IC 348 sources showing signs of
disk activity are associated with our group A objects.

As examples and for comparison with RCW 34 we show in Fig. \ref{fig:orionic348}
the two-colour diagrams for the Orion Nebula Cluster (ONC), using the data of
\citet{hillenbrand1998}, and IC 348, using the data of \citet{luhman2003}. It is
seen that while the ONC has quite a number of objects in region A (which we will
later use), on the other hand IC 348 has very few. None of the two, however, has
the same clustering as seen in RCW 34.

Considering the above, a preliminary conclusion is that the group A objects are
most likely lower mass pre-main sequence stars and that some of the group B
objects, although not having an infrared excess, might also be T Tauri stars.
 
 \begin{figure}[ht]
   \centering \includegraphics[width=7cm,angle=0,clip]{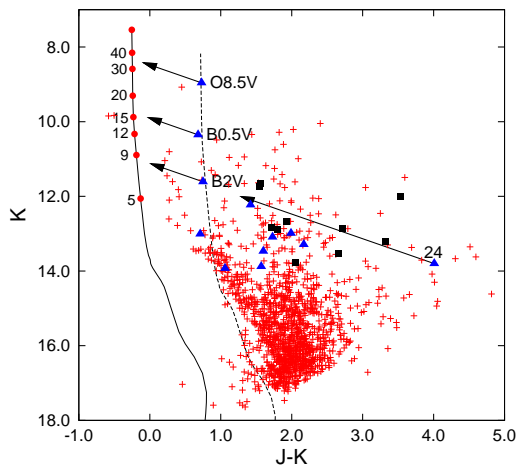}
      \caption{Colour-magnitude diagram for RCW 34. The red crosses are the
        objects detected with the IRSF. The solid black line is the unreddened
        main sequence for a distance of 2.5 kpc. The dashed line is the
        corresponding main-sequence reddened with A$_V$ = 5.1 magnitudes.  Blue
        triangles and solid squares are respectively the late-type main sequence
        stars and pre-main sequence stars from \citet{bik2010}. }
\label{fig:rcw34cmd}
   \end{figure}

\subsection{The colour-magnitude diagram.}
In Fig. \ref{fig:rcw34cmd} we show the K vs J-K colour-magnitude diagram (CMD)
for the 1283 objects detected in J, H, and K with the IRSF (red crosses). We use
the JK CMD in order to compare the positions of the spectroscopically identified
late-type and pre-main sequence stars of \citet{bik2010} (taken from their Table
6) with our NIR detections. The main-sequence (solid line) was taken from the
Padova models \citep{marigo2008} and has been adjusted for distance only using a
distance of 2.5 kpc to RCW 34. The red dots on the main sequence are the
positions of the more massive main sequence stars with their masses (in solar
masses) given to the left of each point. We also show in Fig. \ref{fig:rcw34cmd}
the positions of those stars for which \citet{bik2010} were able to determine
spectral types from the SINFONI observations. The blue triangles are the main
sequence stars and the solid black squares the late-type pre-main sequence
stars. Although there is some overlap in the J-K colours of the two groups, it
is seen that, except for one case, the pre-main sequence stars are on average
redder than the main-sequence stars. The reddest object in the list of
\citet{bik2010} is star 24, with J-K $\sim$ 4. It is classified as an early
K-type dwarf by \citet{bik2010} but of uncertain luminosity class. Its position
is also shown in Fig. \ref{fig:rcw34cmd}.

It is seen that the objects detected with the IRSF populate a very specific area
on the CMD. Most of the objects have $\mathrm{K < 14}$ and $1 < \mathrm{J-K} <
3$ with a clustering around K $\lesssim$ 16 and J-K $\sim$ 2. The effect of the
limited sensitivity of our imaging is also clearly visible as the sharp cutoff
on the lower right hand side of the region covered by the IRSF
objects. Inspection of the equivalent CMD of \citet{bik2010} (their Figure 7)
shows that these authors also detected some objects that lie between K
magnitudes 14 and 16 and for J-K just less than 2 there is a clustering of
objects to the right of the main sequence. Our imaging also shows a large number
of objects at that position on the colour-magnitude diagram. 

Using the visual extinctions calculated by \citet{bik2010} (their Table 4) we
also calculated the dereddened positions of the three early-type stars as well
as that of star 24. The dereddening vectors are shown by the arrows starting at
each of the stars. For the three early-type stars the dereddended positions do
not fall exactly on the main sequence but slightly to the right. Adding a
further approximately 0.9 visual magnitudes will bring the three dereddened
early-type stars on the main sequence. However, we noted that there is a
difference in extinction for the three OB stars as given in Tables 4 and 6 of
\citet{bik2010}. For example for star 1, the exciting star, Table 4 gives an
extinction of 4.2 while it is given as 4.8 in Table 6. The reason for this not
clear. We also note that the extinction toward region II of \citet{bik2010} (see
their Figure 4), which is close to the location of the three OB stars, is A$_v$
= 5.1 magnitudes.  The required visual extinction of 5.1 magnitudes to bring the
unreddened main sequence to the positions of the three OB stars is therefore in
general agreement with the measurements of \citet{bik2010}. It is furthermore
interesting to note that star 24, with an estimated extinction of $15.9 \pm 2.8$
cannot be dereddened at all to the main sequence and if it could, it would not
be a K-dwarf. Star 24 is therefore most likely still in the pre-main sequence
phase.

Taking the three OB stars' positions to indicate the position of the reddened
main sequnce, we can redden the unreddened main sequence accordingly. This is
shown as the dashed line in Fig. \ref{fig:rcw34cmd}. It is seen that the
majority of our IRSF sources still lie significantly to the right of the main
sequence as will be expected for pre-main sequence stars. This result is in
support of our hypothesis that the sample of 1283 objects detected in J, H, and
K contains a significant number of pre-main sequence stars. Those objects lying
significantly to the left of the reddened main sequence are most probably
foreground stars. 

\subsection{The luminosity function}

The luminosity function is an important characteristic of the population of
pre-main sequence stars since it contains information about the age, initial
mass function, and star formation history of that particular population. In
constructing the luminosity function and estimating the ages of members of the
population the ideal would be to know the effective temperature and luminosity
of each individual star from spectroscopic measurements and to then construct a
luminosity-temperature diagram on which model isochrones can overlaid to
estimate the pre-main sequence ages. In the present case we only have NIR
photometry data which is not suitable to determine stellar effective
temperatures due to contamination from excess emission from circumstellar
material. Uncertainties in the photometry can also lead to erroneous conclusions
about the ages of individual objects (see eg \citet{preibisch2012} for an
extensive discussion).

However, as argued by \citet{bontemps2001}, the $J$-band flux is least affected
by contamination from circumstellar emission and is close to the peak of the
photospheric spectral energy distribution for these cool stars and may be used
to {\it estimate} stellar luminosities.  We therefore followed these authors and
estimated the stellar luminosity directly from the absolute $J$-band magnitude
using the relation \citep{bontemps2001}
\begin{equation}
\log_{10}(L_\star) = 1.49 - 0.466\,M_J
\label{eq:lmj}
\end{equation}
The estimated uncertainty on $\log_{10}(L_\star)$ using Eq. \ref{eq:lmj} is 0.19
dex.

\begin{figure}[ht]
   \centering \includegraphics[width=7cm,clip]{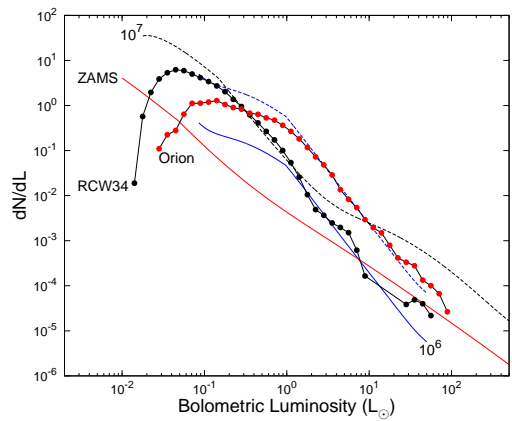}
      \caption
{ Bolometric luminosity distribution for a subset 745 infra-red excess
  sources(black dots). The blue, dashed black, and red lines are the single
  stellar population luminosity functions respectively at $10^6$ and $10^7$
  years and for the ZAMS as calculated from the mass-luminosity relations at
  these times as given by the models of \citet{siess2000}. The luminosity
  function for Orion is given by the black line with red dots.}
\label{fig:lumdists}
\end{figure}

Application of Eq. \ref{eq:lmj} requires calculation of the absolute $J$-band
magnitude and therefore dereddening of the observed $J$-band magnitudes. The
assumption was therefore made that all the group A objects are reddened
CTTS. However, we selected only the subset of 745 group A objects that lie to
the right of reddening line R$_3$, above the T-Tauri locus and with $0.53 <
\mathrm{H-K} < 1$. These objects all have an infrared excess and cannot be
dereddened to the main sequence. Since we do not know the true unreddened
intrinsic colours, each object was dereddened to a random position between the
upper and lower boundaries of the T Tauri locus but still lying on its
individual dereddening line. One instance of dereddening all 745 objects then
results in a luminosity function for the given random positions around the T
Tauri locus. By repeating the procedure a large number of times it is possible
to construct an average luminosity function that is representative of a large
number of realizations of intrinsic colours for the dereddened objects.  All
objects were assumed to lie at a distance of 2.5 kpc to calculate the absolute
magnitudes and therefore the stellar luminosities.

 In Fig. \ref{fig:lumdists} we show with the black solid line the luminosity
 function averaged over 1000 dereddening realizations and normalized such that
 $\int (dN/dL) dL = 1$. Normalization is necessary to be able to compare the
 luminosity functions of different star forming regions as we will do below.
 The distribution is seen to have a peak at about 0.04$\mathrm{L_\odot}$ below
 which there is a sharp cut-off which most likely is a sensitivity
 effect. Except for the last three highest luminosities, the luminosity function
 follows a power law with an index of $-2.34 \pm 0.07$ for luminosities above
 about 0.3 $\mathrm{L_\odot}$. 

The observed luminosity distribution is a result of the combination of the
evolution of the pre-main sequence mass-luminosity relation with time, the
underlying initial mass function and the star formation history of the
region. We attempted to interpret the observed luminosity distribution with
single stellar population (SSP) luminosity functions constructed by using the
mass-luminosity relation for pre-main sequence isochrones at $10^6$ and $10^7$
years and the ZAMS from the \citet{siess2000} models combined with a Kroupa
stellar initial mass function \citep{kroupa2001} for masses between 0.3
$\mathrm{M_\odot}$ and 2 $\mathrm{M_\odot}$. For an SSP the luminosity function,
$dN/dL$, is related to the initial mass function, $dN/dm$, as $dN/dL =
(dN/dm)\times(dm/dL)$. To calculate $dm/dL$ we approximated the mass-luminosity
relation, as given by the \citet{siess2000} model for a specific isochrone, on a
log-log scale with a polinomial which is easily differentiable. The use of the
$10^6$ and $10^7$ year luminosity functions as comparison with the observed
luminosity function should not be interpreted as attemps for absolute age
estimates but as guides to interpret the observed luminosity function. 
The result of this calculation is shown in Fig. \ref{fig:lumdists} where the
luminosity functions at $10^6$ (blue solid line) and $10^7$ (dashed black line)
years have been adjusted vertically to more or less coincide with parts of the
observed luminosity function of RCW 34. The luminosity function for the ZAMS
(solid red line) has been placed at an arbitrary position.  

Inspection of Fig. \ref{fig:lumdists} shows that although the observed
luminosity function can be fitted by a single power law for $\mathrm{L >
  0.3\,L_\odot}$, neither of the $10^6$ or $10^7$ year SSP luminosity functions
can explain the entire observed luminosity function. It is seen that the
theoretical $10^6$ year luminosity function explains the observed luminosity
function rather well for luminosities greater than about $\mathrm{1.4
  \,L_\odot}$.  Below $\mathrm{\sim 1.4 \,L_\odot}$ the theoretical luminosity
dips below the observed luminosity function, suggesting the presence of an older
component. The theoretical luminosity function for an SSP with an age of $10^7$
years is seen to follow the observed luminosity function quite well between
$\mathrm{\sim 0.2 \,L_\odot}$ and just less than $\mathrm{1.4 \,L_\odot}$. For
$\mathrm{L \lesssim 0.2 \,L_\odot}$ the observed luminosity function starts to
bend away from and fall below the theoretical $10^7$ year luminosity
function. This behaviour is most certainly due to the limited sensitivity of our
imaging. Finally we note that the observed luminosity function nowhere has the
behaviour of the luminosity function for the ZAMS suggesting that the group A
objects used here are not main sequence objects.

The fact that a significant part of the RCW 34 luminosity function can be
explained by a combination of the $10^6$ and $10^7$ year SSP luminosity
functions is suggestive that star formation in RCW 34 has been an ongoing
process for about $10^7$ years. As already noted, based on the presence of the
O8.5V star that powers the \ion{H}{2} region, \citet{bik2010} put an upper limit
of $8 \times 10^6 \sim 10^7$ years on the age of RCW 34. The presence of a class
II methanol maser indicates a much more recent episode of star formation
activity associated with RCW 34 since these masers are known to be exclusively
associated with a very early phase of massive star formation
\citep{ellingsen2006}. Furthermore, the 6.7 GHz methanol maser lifetime is
estimated to be between $2.5\times 10^4$ and $4.0\times 10^4$ years
\citep{vanderwalt2005}. From their set of 26 spectroscopically identified
pre-main sequence stars \citet{bik2010} estimate the age of the youngest
pre-main sequence objects to be $(2\pm 1) \times 10^6$ years. Our analysis of
the NIR photometric data of a completely different sample of objects in RCW 34
is therefore in general agreement with that of \citet{bik2010}.

In Fig. \ref{fig:lumdists} we also show the luminosity function for
204 pre-main sequence stars from the Orion Nebular Cluster using the
data of \citet{hillenbrand1998}. The JHK colours of all the objects in
the catalog of \citet{hillenbrand1998} were transformed from the CIT
to the 2MASS system and afterwards subjected to the same selection
criteria that was used to select the subset of 745 group A sources
from which the RCW 34 luminosity function constructed. Dereddening was
done in the same way as for the RCW 34 sources and an average
luminosity function was calculated over 1000 dereddening realizations.

For the Orion sources the luminosity function peaks at about 0.1
$\mathrm{L_\odot}$ compared to 0.04 $\mathrm{L_\odot}$ for RCW 34.  The Orion
luminosity function crosses the RCW 34 luminosity function between 0.2 and 0.3
$\mathrm{L_\odot}$. For L $\gtrsim$ 2 $\mathrm{L_\odot}$ the two luminosity functions
runs almost parallel to each other except for the last two points for RCW 34. In
fact, the main trend of the ONC luminosity function for $\mathrm{L >
  2\,L_\odot}$ is a power law with index -2.17 $\pm$ 0.05 which is very similar
to that of RCW 34. It is also seen that for $\mathrm{L \ge 2~L_\odot}$ the
$10^6$ year SSP luminosity function (dashed blue line) describes the Orion
luminosity function quite well. This is in agreement with independent age
estimates of about $2\times 10^6$ years for the ONC \citep[see
  eg.][]{reggiani2011}. Comparison of the $10^6$ and $10^7$ year SSP luminosity
functions suggest that the slope of the luminosity function for $\mathrm{L >
  2\,L_\odot}$ is definitely dependent on the age of the system. Thus, just by
comparison of the luminosity function of RCW 34 with that of the ONC already
suggests the existence of component with an age of about 1 - 2$\times 10^6$
years in RCW 34.

\subsection{Spatial distribution of objects.}
\begin{figure}[ht]
   \centering \includegraphics[width=6.0cm,angle=0,clip]{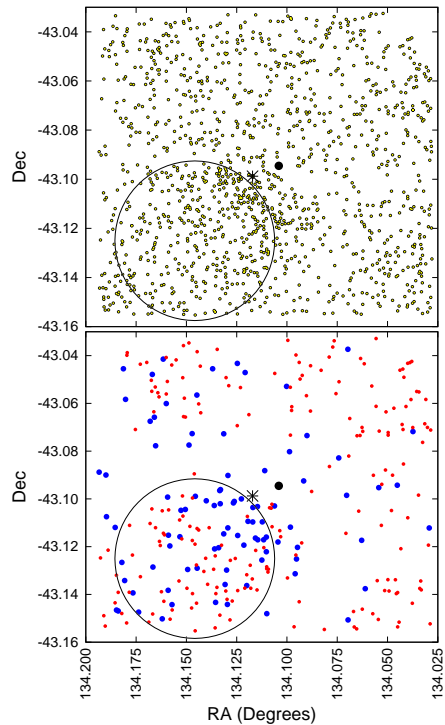}
      \caption{{\it Top panel:} The spatial distribution of all 1283
        objects. {\it Bottom panel:} Spatial distribution for 91 objects with
        $\mathrm{L > 0.5\,L_\odot}$ (blue dots) and 209 objects with $\mathrm{L
          < 0.075\,L_\odot}$ (red dots). The black dot indicates the position of
        the 6.7 GHz methanol maser and the star the position of the O8.5V
        star. The circle in the top and bottom panels indicates the approximate
        size and location of the opacity hole (bubble) as revealed in the dust
        emission of the region \citep[See eg. Figs. 1 \& 3 of][]{bik2010}. Note
        that the coordinates are in decimal degree.}
\label{fig:rcw34dist}
   \end{figure}

 In the top panel of Fig. \ref{fig:rcw34dist} we show the spatial distribution
 of the 1283 objects detected in J, H and K.  Some degree of clustering can be seen in a short band running
 north-east to south-west just south of the central O8.5V star. Otherwise, the
 detected objects seems to be more or less uniformly distributed over the field.

 \begin{figure}[h]
   \centering \includegraphics[width=7.5cm,angle=0,clip]{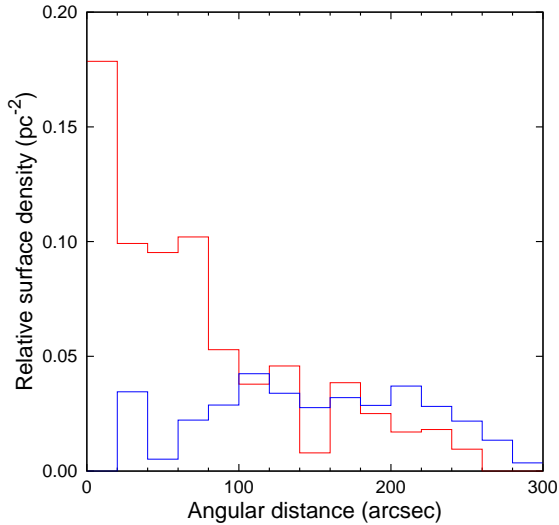}
      \caption{Radial surface density distribution of group A objects with
        luminosity greater than 0.5$\mathrm{L_\odot}$ (red line) and objects
        with luminosity less than 0.075$\mathrm{L_\odot}$ (black line). The
        origin was taken as the position of the O8.5V star.}
\label{fig:agedisthist}
   \end{figure}

Given that our analysis of the luminosity function suggests that star formation
in RCW 34 took place over an extensive period of time, the question is whether
there is any difference in the spacial distribution of the younger and older
pre-main sequence stars.  Here we follow the discussion in the previous section
and use luminosity as an approximate indicator of age with younger objects
having greater luminosity than older ones. To ensure that we have a
statistically sufficient number of objects we select the brighter objects as
those with $\mathrm{L > 0.5\,L_\odot }$. This resulted in 91 objects. For the
older group we selected the 209 objects with $\mathrm{L < 0.075\,L_\odot}$. The
spatial distribution of the two groups is shown in the bottom panel of
Fig. \ref{fig:rcw34dist}

Inspection of the bottom panel of Fig. \ref{fig:rcw34dist} suggests a clustering of
the higher luminosity objects (younger ones; blue dots) closer to the O8.5V star
than the fainter objects (red dots) and in particular in the region of the
opacity hole. To quantify the distributions we calculated the relative surface
density of younger and older objects as a function of distance from the O8.5V
star using a number of annulli.  The two distributions are shown in
Fig. \ref{fig:agedisthist}. The brighter objects clearly shows a significantly
higher probability of being found closer to the O8.5V star than the fainter
objects which have a flatter surface density distribution. In both cases the
decrease in the surface density beyond $\sim$240 arcsec is artificial and due to
the fact that we have not corrected the larger annulli for the smaller surface
area enclosed inside the frame borders.

\citet{reggiani2011} recently also investigated the variation of the surface
density distribution of pre-main sequence stars with age in the
ONC. Interestingly enough these authors found that the older objects have a less
concentrated distribution compared to the younger objects. Qualitatively the
same behaviour is seen in RCW 34. However, it should be kept in mind that the
older lower mass stars are most probably of the same age as the three OB
stars. The concentration of the younger group closer to the three OB stars seems
to suggest that the more recent episode of star formation may have been
triggered by the presence of these three stars. 

\section{Summary and Conclusions}
We presented NIR imaging data on RCW 34 for a $7.8 \times 7.8$ arcmin$^2$ region
centered on the 6.7 GHz methanol maser associated with RCW 34. A total of 1283
objects were detected in J, H, and K-bands. The distribution of these objects on
the two-colour diagram shows a concentration of more than 750 objects for which
the colours are the same as that of confirmed classical T Tauri stars found in
other star forming regions. Given that the position of the main sequence on the
K {\it vs} J-K CMD is determined by the positions of the three OB stars, the
distribution of the IRSF sources on the CMD is also suggestive that a
significant number of these are lower mass pre-main sequence stars.

We also constructed the bolometric luminosity function for the 745 objects and
showed that parts of the luminosity function can be explained by SSP luminosity
functions with ages of $10^6$ and $10^7$ years. The presence of a young
component in the stellar population of RCW 34 is in agreement with the results
of \citet{bik2010} based on 26 spectroscopically identified pre-main sequence
stars as well as the presence of a class II methanol maser which indicate a very
recent episode of massive star formation. Our estimate of $10^7$ years for the
age of an older component is qualitatively in agreement with the main sequence
life times of the three OB stars associated with RCW 34.

Whereas previous studies of RCW 34 focused more on the central region around the
ionizing star, our NIR imaging revealed a more spread out population of low mass
pre-main sequence stars. The younger stars appear to be more concentrated in the
central region around the O8.5V star while the older pre-main sequence stars
seems to be more spread out. 

Considering our NIR results as well as that of \citet{bik2010}, RCW 34 seems to
be a much more interesting star forming region than was perhaps previously
thought. First, there appears to be a very large number of low mass stars formed
over a period of about $10^7$ years with the older being more spread out than
the younger component. Second, it seems rather certain that massive star
formation in RCW 34 took place in two separate events with only three OB stars
forming in the first event and most probably only a single massive star, as
evidenced by the 6.7 GHz maser, forming in the most recent event. The fact that
the brighter (younger) lower mass pre-main sequence stars seems to cluster
around the position of the three OB stars strongly suggests that that the
physical effects these three stars had on the remainder of the molecular cloud
from which they formed could have played a role in the more recent episode of
star formation. Obviously our photometric study need to be followed up by a
spectroscopic study which we plan to do. Apart from verifying our photometric
identifications, will such a study be very usefull to also try to unravel the
star formation history, especially of the low mass stars, in RCW 34. Given that
there also seems to be a significant number of low mass pre-main sequence stars
spread out almost uniformly over the $7.8 \times 7.8$ arcmin$^2$ region might
require a different scenario for the star formation history in RCW 34 than that
suggested by \citet{bik2010}. Although RCW 34 cannot be regarded as peculiar
anymore, it is seems to be interesting and sufficiently different from other
star forming regions to require further investigation.

\acknowledgments We would like to thank an anonomous referee for constructive
comments to improve the paper. This work was supported by the National Research
Foundation under Grant number 2053475.  

\bibliographystyle{apj}
\bibliography{ref}

\begin{thebibliography}{38}
\expandafter\ifx\csname natexlab\endcsname\relax\def\natexlab#1{#1}\fi

\bibitem[{{Balog} {et~al.}(2004){Balog}, {Kenyon}, {Lada}, {Barsony},
  {Vink{\'o}}, \& {G{\'a}spa{\'r}}}]{balog2004}
{Balog}, Z., {Kenyon}, S.~J., {Lada}, E.~A., {Barsony}, M., {Vink{\'o}}, J., \&
  {G{\'a}spa{\'r}}, A. 2004, \aj, 128, 2942

\bibitem[{{Barentsen} {et~al.}(2011){Barentsen}, {Vink}, {Drew}, {Greimel},
  {Wright}, {Drake}, {Martin}, {Valdivielso}, \& {Corradi}}]{barentsen2011}
{Barentsen}, G., {et~al.} 2011, \mnras, 415, 103

\bibitem[{{Bik} {et~al.}(2010){Bik}, {Puga}, {Waters}, {Horrobin}, {Henning},
  {Vasyunina}, {Beuther}, {Linz}, {Kaper}, {van den Ancker}, {Lenorzer},
  {Churchwell}, {Kurtz}, {Kouwenhoven}, {Stolte}, {de Koter}, {Thi},
  {Comer{\'o}n}, \& {Waelkens}}]{bik2010}
{Bik}, A., {et~al.} 2010, \apj, 713, 883

\bibitem[{{Bontemps} {et~al.}(2001){Bontemps}, {Andr{\'e}}, {Kaas}, {Nordh},
  {Olofsson}, {Huldtgren}, {Abergel}, {Blommaert}, {Boulanger}, {Burgdorf},
  {Cesarsky}, {Cesarsky}, {Copet}, {Davies}, {Falgarone}, {Lagache},
  {Montmerle}, {P{\'e}rault}, {Persi}, {Prusti}, {Puget}, \&
  {Sibille}}]{bontemps2001}
{Bontemps}, S., {et~al.} 2001, \aap, 372, 173

\bibitem[{{Bowler} {et~al.}(2009){Bowler}, {Waller}, {Megeath}, {Patten}, \&
  {Tamura}}]{bowler2009}
{Bowler}, B.~P., {Waller}, W.~H., {Megeath}, S.~T., {Patten}, B.~M., \&
  {Tamura}, M. 2009, \aj, 137, 3685

\bibitem[{{Braz} \& {Scalise}(1982)}]{braz1982}
{Braz}, M.~A., \& {Scalise}, Jr., E. 1982, \aap, 107, 272

\bibitem[{{Caswell} \& {Haynes}(1987)}]{caswell1987}
{Caswell}, J.~L., \& {Haynes}, R.~F. 1987, \aap, 171, 261

\bibitem[{{Cieza} {et~al.}(2005){Cieza}, {Kessler-Silacci}, {Jaffe}, {Harvey},
  \& {Evans}}]{cieza2005}
{Cieza}, L.~A., {Kessler-Silacci}, J.~E., {Jaffe}, D.~T., {Harvey}, P.~M., \&
  {Evans}, II, N.~J. 2005, \apj, 635, 422

\bibitem[{{Dahm} \& {Simon}(2005)}]{dahm2005}
{Dahm}, S.~E., \& {Simon}, T. 2005, \aj, 129, 829

\bibitem[{{Devine} {et~al.}(2008){Devine}, {Churchwell}, {Indebetouw},
  {Watson}, \& {Crawford}}]{devine2008}
{Devine}, K.~E., {Churchwell}, E.~B., {Indebetouw}, R., {Watson}, C., \&
  {Crawford}, S.~M. 2008, \aj, 135, 2095

\bibitem[{{Ellingsen}(2006)}]{ellingsen2006}
{Ellingsen}, S.~P. 2006, \apj, 638, 241

\bibitem[{{Fazio} {et~al.}(2004){Fazio}, {Hora}, {Allen}, {Ashby}, {Barmby},
  {Deutsch}, {Huang}, {Kleiner}, {Marengo}, {Megeath}, {Melnick}, {Pahre},
  {Patten}, {Polizotti}, {Smith}, {Taylor}, {Wang}, {Willner}, {Hoffmann},
  {Pipher}, {Forrest}, {McMurty}, {McCreight}, {McKelvey}, {McMurray}, {Koch},
  {Moseley}, {Arendt}, {Mentzell}, {Marx}, {Losch}, {Mayman}, {Eichhorn},
  {Krebs}, {Jhabvala}, {Gezari}, {Fixsen}, {Flores}, {Shakoorzadeh}, {Jungo},
  {Hakun}, {Workman}, {Karpati}, {Kichak}, {Whitley}, {Mann}, {Tollestrup},
  {Eisenhardt}, {Stern}, {Gorjian}, {Bhattacharya}, {Carey}, {Nelson},
  {Glaccum}, {Lacy}, {Lowrance}, {Laine}, {Reach}, {Stauffer}, {Surace},
  {Wilson}, {Wright}, {Hoffman}, {Domingo}, \& {Cohen}}]{fazio2004}
{Fazio}, G.~G., {et~al.} 2004, \apjs, 154, 10

\bibitem[{{Herbst}(1975{\natexlab{a}})}]{herbst1975a}
{Herbst}, W. 1975{\natexlab{a}}, \aj, 80, 212

\bibitem[{{Herbst}(1975{\natexlab{b}})}]{herbst1975b}
---. 1975{\natexlab{b}}, \aj, 80, 683

\bibitem[{{Heydari-Malayeri}(1988)}]{hm1988}
{Heydari-Malayeri}, M. 1988, \aap, 202, 240

\bibitem[{{Hillenbrand} {et~al.}(1998){Hillenbrand}, {Strom}, {Calvet},
  {Merrill}, {Gatley}, {Makidon}, {Meyer}, \& {Skrutskie}}]{hillenbrand1998}
{Hillenbrand}, L.~A., {Strom}, S.~E., {Calvet}, N., {Merrill}, K.~M., {Gatley},
  I., {Makidon}, R.~B., {Meyer}, M.~R., \& {Skrutskie}, M.~F. 1998, \aj, 116,
  1816

\bibitem[{{Koornneef}(1983)}]{koornneef1983}
{Koornneef}, J. 1983, \aaps, 51, 489

\bibitem[{{Kroupa}(2001)}]{kroupa2001}
{Kroupa}, P. 2001, \mnras, 322, 231

\bibitem[{{Lada}(2010)}]{lada2010}
{Lada}, C.~J. 2010, Royal Society of London Philosophical Transactions Series
  A, 368, 713

\bibitem[{{Lada} \& {Adams}(1992)}]{lada1992}
{Lada}, C.~J., \& {Adams}, F.~C. 1992, \apj, 393, 278

\bibitem[{{Lada} {et~al.}(1996){Lada}, {Alves}, \& {Lada}}]{lada1996}
{Lada}, C.~J., {Alves}, J., \& {Lada}, E.~A. 1996, \aj, 111, 1964

\bibitem[{{Lada} \& {Lada}(1995)}]{ladaea1995}
{Lada}, E.~A., \& {Lada}, C.~J. 1995, \aj, 109, 1682

\bibitem[{{Li} {et~al.}(1997){Li}, {Evans}, \& {Lada}}]{li1997}
{Li}, W., {Evans}, II, N.~J., \& {Lada}, E.~A. 1997, \apj, 488, 277

\bibitem[{{Luhman} {et~al.}(1998){Luhman}, {Rieke}, {Lada}, \&
  {Lada}}]{luhman1998}
{Luhman}, K.~L., {Rieke}, G.~H., {Lada}, C.~J., \& {Lada}, E.~A. 1998, \apj,
  508, 347

\bibitem[{{Luhman} {et~al.}(2003){Luhman}, {Stauffer}, {Muench}, {Rieke},
  {Lada}, {Bouvier}, \& {Lada}}]{luhman2003}
{Luhman}, K.~L., {Stauffer}, J.~R., {Muench}, A.~A., {Rieke}, G.~H., {Lada},
  E.~A., {Bouvier}, J., \& {Lada}, C.~J. 2003, \apj, 593, 1093

\bibitem[{{Marigo} {et~al.}(2008){Marigo}, {Girardi}, {Bressan}, {Groenewegen},
  {Silva}, \& {Granato}}]{marigo2008}
{Marigo}, P., {Girardi}, L., {Bressan}, A., {Groenewegen}, M.~A.~T., {Silva},
  L., \& {Granato}, G.~L. 2008, \aap, 482, 883

\bibitem[{{Meyer} {et~al.}(1997){Meyer}, {Calvet}, \&
  {Hillenbrand}}]{meyer1997}
{Meyer}, M.~R., {Calvet}, N., \& {Hillenbrand}, L.~A. 1997, \aj, 114, 288

\bibitem[{{Preibisch}(2012)}]{preibisch2012}
{Preibisch}, T. 2012, Research in Astronomy and Astrophysics, 12, 1

\bibitem[{{Reggiani} {et~al.}(2011){Reggiani}, {Robberto}, {da Rio}, {Meyer},
  {Soderblom}, \& {Ricci}}]{reggiani2011}
{Reggiani}, M., {Robberto}, M., {da Rio}, N., {Meyer}, M.~R., {Soderblom},
  D.~R., \& {Ricci}, L. 2011, \aap, 534, A83

\bibitem[{{Rieke} \& {Lebofsky}(1985)}]{rieke1985}
{Rieke}, G.~H., \& {Lebofsky}, M.~J. 1985, \apj, 288, 618

\bibitem[{{Rodgers} {et~al.}(1960){Rodgers}, {Campbell}, \&
  {Whiteoak}}]{rodgers1960}
{Rodgers}, A.~W., {Campbell}, C.~T., \& {Whiteoak}, J.~B. 1960, \mnras, 121,
  103

\bibitem[{{Siess} {et~al.}(2000){Siess}, {Dufour}, \& {Forestini}}]{siess2000}
{Siess}, L., {Dufour}, E., \& {Forestini}, M. 2000, \aap, 358, 593

\bibitem[{{Strom} {et~al.}(1989){Strom}, {Strom}, {Edwards}, {Cabrit}, \&
  {Skrutskie}}]{strom1989}
{Strom}, K.~M., {Strom}, S.~E., {Edwards}, S., {Cabrit}, S., \& {Skrutskie},
  M.~F. 1989, \aj, 97, 1451

\bibitem[{{Teixeira} {et~al.}(2004){Teixeira}, {Fernandes}, {Alves}, {Correia},
  {Santos}, {Lada}, \& {Lada}}]{teixeira2004}
{Teixeira}, P.~S., {Fernandes}, S.~R., {Alves}, J.~F., {Correia}, J.~C.,
  {Santos}, F.~D., {Lada}, E.~A., \& {Lada}, C.~J. 2004, \aap, 413, L1

\bibitem[{{van den Bergh} \& {Herbst}(1975)}]{vandenbergh1975}
{van den Bergh}, S., \& {Herbst}, W. 1975, \aj, 80, 208

\bibitem[{{van der Walt}(2005)}]{vanderwalt2005}
{van der Walt}, J. 2005, \mnras, 360, 153

\bibitem[{{Vittone} {et~al.}(1987){Vittone}, {de Martino}, {Giovannelli}, \&
  {Rossi}}]{vittone1987}
{Vittone}, A.~A., {de Martino}, D., {Giovannelli}, F., \& {Rossi}, C. 1987,
  \aap, 179, 157

\bibitem[{{Wouterloot} \& {Brand}(1989)}]{wouterloot1989}
{Wouterloot}, J.~G.~A., \& {Brand}, J. 1989, \aaps, 80, 149

\end{thebibliography}

\end{document}